\documentclass{ws-ijmpe}

\def\eV{\hbox{ eV}}

\def\GeV{\hbox{ GeV}}

\begin{document}


\markboth
{M. G\'o\'zd\'z and M. Rogatko} 
{Neutrino Oscillations in Strong Gravitational Fields}

%
%

\title{NEUTRINO OSCILLATIONS\\IN STRONG GRAVITATIONAL FIELDS}

\author{MAREK G\'O\'ZD\'Z$^1$ and MAREK ROGATKO$^2$} 
\address{
$^1$Department of Informatics, $^2$Department of Physics,\\
Maria Curie-Sk{\l}odowska University, \\
pl. Marii Curie--Sk{\l}odowskiej 5, 20-031 Lublin, Poland \\
mgozdz@kft.umcs.lublin.pl, rogat@kft.umcs.lublin.pl}

\maketitle

\begin{history}                 %
\end{history}                  	%


\begin{abstract}
  Neutrinos do oscillate, which up to our best knowledge implies that
  they are massive particles. As such, neutrinos should interact with
  gravitational fields. As their masses are tiny, the gravitational
  fields must be extremely strong. In this paper we study the influence
  of black holes described by non-trivial topologies on the neutrino
  oscillations. We present approximate analytical and numerical
  solutions of certain specific cases.
\end{abstract}


\section{Introduction}

Neutrino oscillations are one of the most interesting phenomena in
particle physics. They were anticipated long time ago,\cite{neutosc} but
their detection was complicated due to very weak interaction of these
particles. Nowadays, these phenomena have been detected and studied in
the case of solar neutrinos, reactor neutrinos, and atmospheric
interaction of cosmic rays.\cite{nu-osc,osc} Up to our best knowledge
massless particles cannot oscillate, and so neutrinos (at least two out
of three) must have mass, which is estimated from the supernova and
other astrophysical observations to be roughly 0.3~eV.\cite{astro}

In this paper we are going to discuss the (very weak) interaction
between massive neutrinos and a~gravitational field. We focus here on
the change of oscillation rate for neutrinos propagating close to black
holes. In principle, the strong gravitational field should modify the
vacuum oscillation results, introducing additional phase shift. The
quantum mechanical phase of neutrinos in the Schwarzschild spacetime was
presented in Ref.\cite{car97}, where the authors discussed propagation
including the possible effect of interaction with matter (MSW
effect). The non-radial propagation of neutrinos in the aforementioned
background was elaborated in Ref.\cite{for97}, while the critical
examination of the gravitationally induced quantum mechanical phases in
neutrino oscillations was given in Ref.\cite{bha99}.

We have found the results of Ref.\cite{for97} particularly interesting,
as the expressions for the oscillation phase $\Phi_k$, gained by
neutrinos during propagation in a~gravitational field, contained terms
proportional to $M$, the mass of the gravitating source. This
observation suggests that for certain astrophysical objects, like
super-massive black holes for example, the contribution to $\Phi_k$ may
be substantial. By performing a~more carefull analysis we show that
unfortunatelly this is not the case.

In our work we use a~general form of the metric, which allows us to
discuss not only the flat Schwarzschild background, but also the case of
certain topological defects (like a~black hole pierced by a~cosmic
string and a~black hole-global monopole system).

\section{Neutrino oscillations in the vicinity of a black hole}

In this section we derive the formula for the quantum mechanical phase,
acquired by a neutrino which propagates in a strong gravitational field,
like in the vicinity of a black hole. Let us start with the following
line element,
\begin{equation}
  \label{eq:ds1}
  ds^2 = -B(r) dt^2 + \frac{dr^2}{B(r)} + \tilde C(r) r^2 d\theta^2 + C(r)
  r^2\sin^2\theta d\phi^2,
\end{equation}
where $B(r)$, $\tilde C(r)$, and $C(r)$ are functions of the radial
coordinate $r$. In our case we are interested in the motion of
a~neutrino in the space-time around a~black hole described by
Eq.~(\ref{eq:ds1}). Because of the spherical symmetry we may always
confine this motion to the plane $\theta=\pi/2$. This simplifies the
line element to the following form:
\begin{equation}
  \label{eq:ds2}
  ds^2 = -B(r) dt^2 + \frac{dr^2}{B(r)} + C(r) r^2 d\phi^2.
\end{equation}
Moreover, we assume that the coefficient functions $B(r)$ and $C(r)$
do not depend on time nor on the $\phi$ coordinate. It follows that the
canonical momenta $p_t^{(k)}$ and $p_\phi^{(k)}$ will play the role of
constants of motion: the energy $E_k$ and angular momentum $J_k$ of the
neutrino $k$-th mass eigenstate, as seen by an observer at infinity.

The phase, gained by a~neutrino during propagation from a space-time
point $A$ to a~space-time point $B$, written in a~covariant form,
reads\cite{for97,sto79}
\begin{equation}
  \label{eq:phi-def}
  \Phi_k = \int_A^B p_\mu^{(k)} dx^\mu,
\end{equation}
where $p_\mu^{(k)}$ is the four-momentum of the $k$-th mass eigenstate
of the neutrino, characterized by the mass $m_k$, $ p_\mu^{(k)} = m_k
g_{\mu\nu} \frac{dx^\nu}{ds}$. These quantities are related to each other and to the mass $m_k$ by the mass-shell relation $m_k^2 =
g^{\mu\nu} p_\mu^{(k)} p_\nu^{(k)}$.

Following closely the method presented in Ref.\cite{for97} (a~detailed
presentation is beyond the scope of this contribution) we find, that the
metric parameter $C(r)$, describing the possible topological
non-triviality, will contribute only if the relativistic expansion in
the neutrino mass to energy ratio squared parameter, $(m_k^2/E_k^2)$,
will be performed up to the second order. Consequently, after a~rather
lengthy computation, we finish with
\begin{equation}
  \label{eq:phi2}
  \Phi_k = -\int_{r_A}^{r_B} 
  \frac{E_k dr}{\sqrt{1-B(r)\frac{d^2}{C(r)r^2}}} \left [
    \frac{m_k^2}{2E_k^2} - \frac12 \left( \frac{m_k^2}{2E_k^2} \right)^2
    \frac{\frac{d^2}{C(r)r^2} (1+2B(r)) + B(r)}{1-B(r)\frac{d^2}{C(r)r^2}}
  \right],
\end{equation}
where $d$ denotes the impact parameter. We notice that when the
$(m_k^2/E_k^2)^2$ is neglected, the result of Ref.\cite{for97} is
reproduced. Also, the minus sign is irrelevant and we will drop it for
simplicity.

\section{Flavour changing probability in neutrino oscillations}

Knowing the phase gained by neutrinos during their propagation one may
ask about the probability of a~neutrino to change its flavour as
a~function of the distance from the source. Let us for simplicity limit
ourselves to the two-neutrino case. Then the flavour changing
probability in its text-book form is
\begin{equation}
  {\cal P} = \sin^2(2\theta)\sin^2\left(\frac{\Delta E}{2}t\right),
  \label{eq:Pclass}
\end{equation}
$\theta$ being the mixing angle. In our case, the time-dependent phases
have to be modified by the phase coming from the gravitational
field. The latter, however, depends on the impact parameter $d$, which
represents the distance in which the neutrino passes the black
hole. Therefore we recognize two sources of possible interference among
neutrino states:\cite{for97} between different mass eigenstates going
along the same path, and between mass eigenstates going along slightly
different paths. We call the different paths ``long'' (L) and ``short''
(S) and rewrite the standard definitions of the time-dependent fields as
\begin{eqnarray}
  |\nu_e(t)\rangle &=&
  \frac{\cos\theta}{2}
  (e^{-i(E_1t+\Phi_1^L)}+e^{-i(E_1t+\Phi_1^S)}) |\nu_1\rangle \nonumber \\
  &+& \frac{\sin\theta}{2}
  (e^{-i(E_2t+\Phi_2^L)}+e^{-i(E_2t+\Phi_2^S)}) |\nu_2\rangle, \\
  |\nu_\mu(t)\rangle &=&
  -\frac{\sin\theta}{2}
  (e^{-i(E_1t+\Phi_1^L)}+e^{-i(E_1t+\Phi_1^S)}) |\nu_1\rangle \nonumber \\
  &+& \frac{\cos\theta}{2}
  (e^{-i(E_2t+\Phi_2^L)}+e^{-i(E_2t+\Phi_2^S)}) |\nu_2\rangle.
\end{eqnarray}
The usual expression for the probability now turns into 
\begin{eqnarray}
  {\cal P}_{\rm grav}&=& \frac{\sin^2(2\theta)}{8}
  \Big[ 2
  + \cos(\Phi_1^L - \Phi_1^S)
  + \cos(\Phi_2^L - \Phi_2^S) \nonumber\\
  &-& \cos(\Phi_1^L - \Phi_2^L + \Delta Et)
  - \cos(\Phi_1^S - \Phi_2^S + \Delta Et) \nonumber\\
  &-& \cos(\Phi_1^S - \Phi_2^L + \Delta Et)
  - \cos(\Phi_1^L - \Phi_2^S + \Delta Et) \Big],
  \label{eq:Pgrav}
\end{eqnarray}
which represents the probability of $\nu_e\to \nu_\mu$ transition in the
background of the gravitational field. As a~check one notices, that for
$\Phi=0$ the usual probability Eq.~(\ref{eq:Pclass}) is recovered. In
what follows we will attempt to estimate the full probability ${\cal
  P}_{\rm grav}$ as well as the difference ${\cal P}_{\rm grav}-{\cal
  P}$ for some special cases.

\section{Approximate solution for the phase $\Phi_k$}

The integral Eq.~(\ref{eq:phi2}) is not solvable exactly in terms of
elementary or special functions. We will attempt, however, to give an
estimate of the solution for two extreme cases: a~super-massive and
a~micro black hole.

An approach that has been used in Ref.\cite{for97} was the so-called
weak field approximation, in which it is globally assumed that $GM\ll
r$. This has lead to the solution $\Phi_k \sim m_k^2 G M / E_0$, cf.
Eq.~(59) in Ref.\cite{for97}, which increases with the mass of the
source of the gravitational field. As this approximation is valid in
certain cases, we will not use it here. One example for which it cannot
be used is the super-massive black hole which is believed to reside in
the center of our galaxy. Its mass is estimated to be around
$10^{37}$~kg.\cite{milkywaybh} One may easily check that neither for $r$
close to its event horizon $\sim 10^{10}$~m, nor for $r$ being
approximately the distance between the Earth and the center of the Milky
Way $\sim 10^{20}$~m, this approach is not justified.

A~few words about possible metrics describing topological defects are in
order. For example, a~black hole pierced by a~cosmic string is described
by the metric (\ref{eq:ds2}) with $B(r)=1-\frac{R}{r}$, $C(r)=1-4\mu$,
while a~black hole with a~global monopole has $B(r)=1-8\pi
G\eta^2-\frac{R}{r}$, $C(r)=1-8\pi G\eta^2$. Both of these cases,
although physically different, are mathematically equivalent, with
$B(r)$ being a~function of the black hole's mass $M$ and the distance
$r$, and $C(r)=$~const. The parameters $\mu$ and $\eta$ are purely
theoretical and only rough bounds for them can be formulated, but they
are generically very small. In order not to violate existing
observations, $C(r)$ is believed to be of the order $1-10^{-\alpha}$
with $\alpha=6-15$.\cite{defect}

\begin{figure}
  \hbox{
  \hbox{
  \begin{minipage}[l]{0.45\linewidth}
    \includegraphics[width=0.9\textwidth]{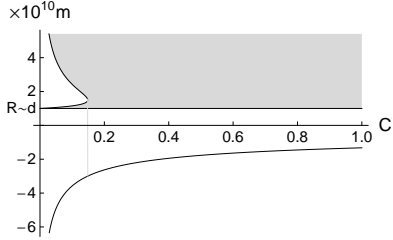}    
  \end{minipage}
  }\hbox{
  \begin{minipage}[l]{0.45\linewidth}
    \caption{\label{fig:1}Distance $r$ as a function of the metric
      parameter $C$, that corresponds to real phases $\Phi_k$ in the
      case of a~super massive black hole from the center of the Milky
      Way. The shaded region represents physically acceptable
      solutions.}
  \end{minipage}
  }}
\end{figure}

For the asymptotic case $r\to\infty$ one may replace both $B(r)$ and
$C(r)$ by 1. This yields in the leading order
\begin{equation}
  \Phi_k^{\rm far} = \frac{m_k^2}{2E_k}\int_{r_A}^{r_B}
  \frac{dr}{\sqrt{1-\frac{d^2}{r^2}}} = \frac{m_k^2}{2E_k} \left[
    r\sqrt{1-\frac{d^2}{r^2}} \right]_{r_A}^{r_B} \approx
  \frac{m_k^2}{2E_k} (r_B-r_A).
\end{equation}
For the close limit, $r\approx d\ge R$, we approximate $B(r)\approx
1-\frac{R}{d}$. This simplification results in
\begin{equation}
  \Phi_k^{\rm close} = \frac{m_k^2}{2E_k} \left[
    r\sqrt{1-\left(1-\frac{R}{d}\right)\frac{d^2}{Cr^2}} \right]_{d}^{d'}.
\end{equation}
One may check by solving the inequality $r^3-d^2 r/C + d^2 R/C >0$, that
if only $r>R$ there is always a~$C$ such that $\Phi_k$ is real. An
example is presented in Fig.~\ref{fig:1}, in which the
Schwarzchild radius has been taken to be $10^{10}$~m. This value
corresponds to the super-heavy ($M\sim 10^{37}$~kg) black hole that is
anticipated to reside in the center of our galaxy. On the other hand,
for a~micro black hole ($M\sim 10^{-27}$~kg) that may appear in the LHC
experiments, Fig.~\ref{fig:1} has to be rescaled such that
$R\sim 10^{-54}$~m.


The generic approximation presented above may be reformulated in some
special numerical cases by taking the leading terms, which dominate
significantly over the others. For instance, in the super massive case,
the close limit up to the second order in relativistic expansion is
given by
\begin{equation}
  \label{eq:sh-close}
  \Phi_k^{\rm close} \approx
  \frac{E_k}{\sqrt{2GMd^2}} \frac{m_k^2}{2E_k^2} 
  \left [ 
    \frac25 \left(1+\frac{m_k^2}{4E_k^2} \right) r^{\frac52} +
    \frac19 \frac{m_k^2}{E_k^2d^2} r^{\frac92}
  \right ]_d^{d'}.
\end{equation}
On the other hand, the close limit for a~micro black hole takes the form
\begin{equation}
  \Phi_k^{\rm close} \approx E_k \frac{m_k^2}{2E_k^2} \Bigg[ \sqrt{r^2-d^2}
  - \frac{1}{2} \frac{m_k^2}{2E_k^2}
  \frac{r^2-5d^2}{\sqrt{r^2-d^2}} \Bigg]_d^{d'}.
  \label{eq:mbh-close}
\end{equation}
The far limits are basically unaffected, as all metrics we may be
interested in are asymptotically flat.

\begin{figure}[b]
  \centering
  \includegraphics[angle=270,width=0.4\textwidth]{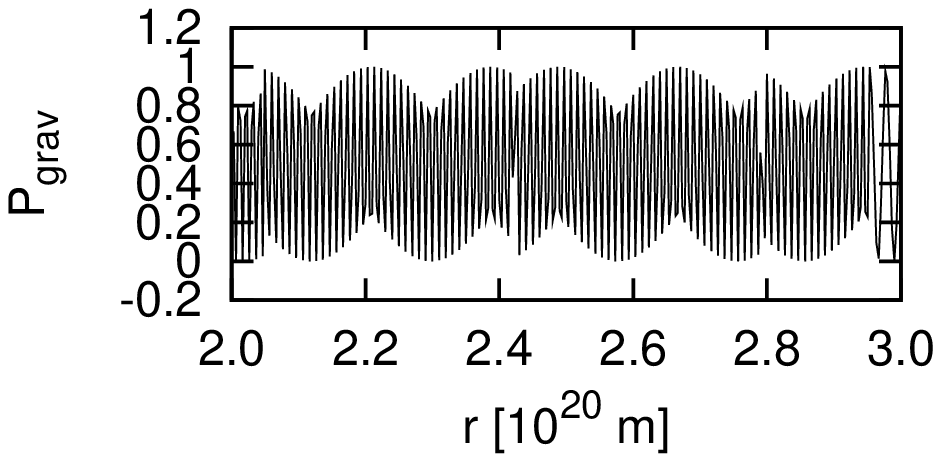}
  \includegraphics[angle=270,width=0.4\textwidth]{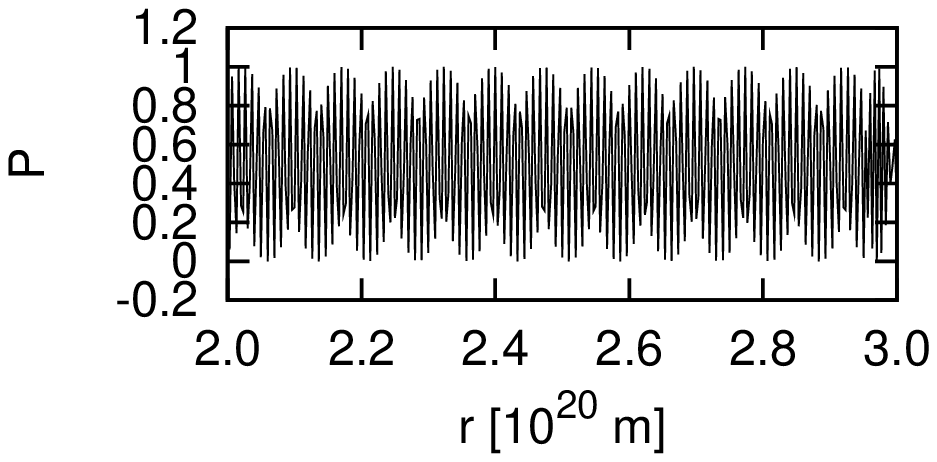}
  \caption{\label{fig:2} Flavour changing probability ${\cal P}_{\rm
      grav}$ in neutrino oscillations as a function of the distance to
    the Milky Way black hole. The result based on the standard formula
    ${\cal P}$ is shown for reference (right panel). Neutrino energy
    $E_k=1\GeV$. See also Fig.~\protect\ref{fig:3}.}
\end{figure}

\begin{figure*}[t]
  \centering
  \includegraphics[angle=270,width=0.8\textwidth]{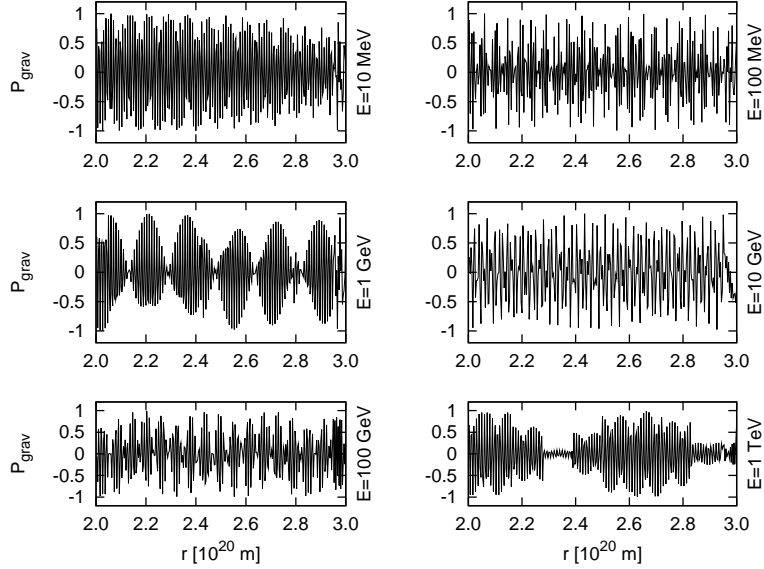}
  \caption{\label{fig:3} Difference ${\cal P}_{\rm grav}-{\cal P}$ for
    various energies $E_k$ of the neutrinos.}
\end{figure*}

The oscillation probabilities for the super-massive black hole are
depicted in Fig.~\ref{fig:2}. We recall here, that the Earth, thus our
would-be observation point, is roughly at the distance $r\approx
2.5\times 10^{20}{\rm m}$. The actual numbers used in the calculations
were $d_S=R$, $d_L=100R$, $m_1=0.30\eV$, $m_2=0.29\eV$, and $\Delta
E=1\eV$. In Fig.~\ref{fig:3} we have collected the functions ${\cal
  P}_{\rm grav}-{\cal P}$ calculated for different energies of the
neutrinos. In all the cases the oscillatory behaviour with interference
patterns is eminent.


\begin{figure}[b]
  \centering
  \includegraphics[angle=270,width=0.4\textwidth]{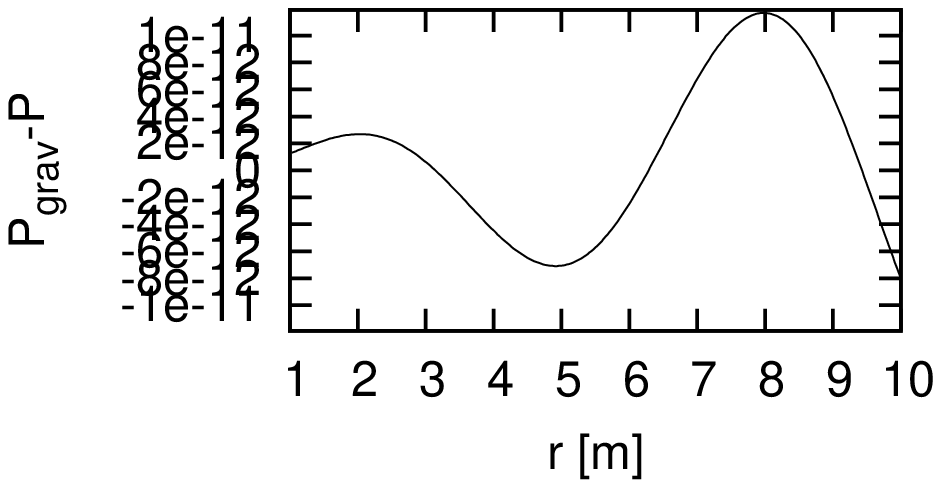}
  \includegraphics[angle=270,width=0.4\textwidth]{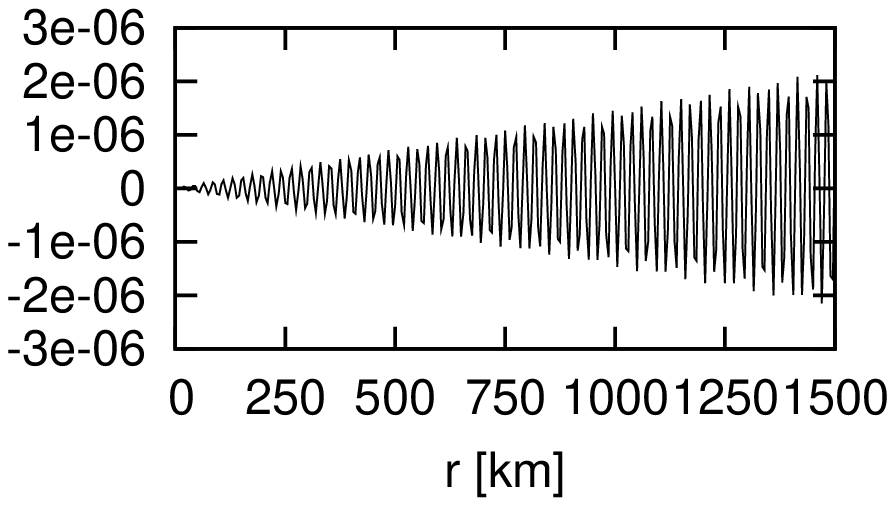}
  \caption{\label{fig:4} Difference ${\cal P}_{\rm grav}-{\cal P}$ as
    a~function of the distance to the micro black hole. Neutrino energy
    $E_k=1\GeV$.}
\end{figure}

Another possible case of interest is a~micro black hole. Given that
certain theories which assume the existence of additional spatial
dimensions are true, at the energy scales which will be reached in the
Large Hadron Collider in CERN some extradimensional black holes should
appear. Their mass is expected to be of the order of $10^{-27}$~kg,
which corresponds to the Schwarzschild radius $R\sim 10^{-54}$~m. These
extremely tiny objects would almost instantly evaporate, however, some
models predict similar objects to be created and travel almost freely
through the Universe. As such, they may act as gravitational lenses (in
the same way as does regular black holes and other massive dark
objects). The results are presented in Fig.~\ref{fig:4} from which one
sees that the difference in transition probabilities may reach the order
of $10^{-6}$ and more on the distance of at least 1500 km. This may
hypothetically give some chances for the future very long baseline
neutrino oscillation experiments to detect the presence of such an
object in the area of the beam.

\section{Conclusions}

Basing on the results presented in Ref.~\cite{for97}, in which the phase
$\Phi_k$ is proportional to $M$, one may expect that the black holes
should generate a~gravitational field strong enough, to give
a~substantial contribution to the neutrino oscillations. The conclusion
from our calculations is quite contrary.

Even though the general formula is known, see Eq.~(\ref{eq:phi2}), its
solution depends on the actual metric parameters, and in most cases
requires pure numerical treatment. We have managed to formulate
approximate analytical solutions for two distinct examples of a
super-massive and a micro black hole with topological defects. Numerical
illustrations of the flavour changing probability as a function of the
distance and neutrino energies have also been presented. In the case of
the super-massive black hole, the interference patterns are quite
different for different energies. It is, however, difficult to imagine
how one would be able to use this knowledge in practice. A more
promising case is the micro black hole. Firstly, using accelerators like
the LHC or even more powerful which will be built in the future, one may
pretty well localize the spot in which such a (hypothetical) black hole
will be produced. The difference in flavour changing probabilities,
$10^{-6}$ for 1500 km up to almost $10^{-5}$ for the Earth diameter, are
still not possible to detect now, but such sensitivity may be probably
reached in the future.

\end{document}